\documentclass[final,5p,times,twocolumn]{elsarticle}

\usepackage{amssymb}
\usepackage{amsmath}

\usepackage{balance}
\usepackage{pgfplots}
\usepackage{amsthm}
\usepackage{url}
\usepackage{epstopdf}
\usepackage{orcidlink}
\usepackage{enumerate}
\usepackage[caption=false]{subfig}
\usepackage{booktabs}

\usepgfplotslibrary{groupplots,dateplot}
\usetikzlibrary{patterns,shapes.arrows}
\pgfplotsset{compat=newest}
\usepackage{dsfont}
\newtheorem{theorem}{Theorem}

\newtheorem{remark}{Remark}
\DeclareMathOperator*{\sgn}{{\rm sgn}}

\DeclareMathOperator*{\argmax}{\arg\!\max}
\newcommand{\revision}[1]{\textcolor{blue}{#1}}
\newcommand{\revisiontwo}[1]{\textcolor{blue}{#1}}
\newcommand{\revisionthree}[1]{\textcolor{blue}{#1}}
\renewcommand{\revision}{}
\renewcommand{\revisiontwo}{}
\renewcommand{\revisionthree}{}

\journal{Information Sciences}

\begin{document}

\begin{frontmatter}



\title{Static and dynamic jamming games over wireless channels with mobile strategic players}

\author[label1]{Giovanni Perin\,\orcidlink{0000-0002-7333-3004}\corref{cor1}}
\affiliation[label1]{organization={Department of Information Engineering, University of Brescia},
             addressline={via Branze 38},
             city={Brescia},
             postcode={25123},
             country={Italy}}

\affiliation[label2]{organization={Department of Information Engineering, University of Padova},
             addressline={via Gradenigo 6/b},
             city={Padova},
             postcode={35131},
             country={Italy}}
\cortext[cor1]{Corresponding author, email: giovanni.perin@unibs.it}

\author[label2]{Leonardo Badia\,\orcidlink{0000-0001-5770-1199}} 


\begin{abstract}
We study a wireless jamming problem consisting of the competition between a legitimate receiver and a jammer, \revisionthree{modeled as} a zero-sum game where the value to maximize/minimize is the channel capacity at the receiver's side. \revisionthree{Most approaches} in the literature consider the two players to be stationary nodes. Instead, we investigate what happens when they can change location, specifically moving along a linear geometry. We frame this at first as a static game, which can be solved in closed form, and subsequently we extend it to a dynamic game under three different versions \revisionthree{concerning} completeness/perfection of mutual information about the adversary's position, corresponding to different assumptions of concealment/sequentiality of the moves, respectively. We first provide some theoretical conditions that hold for the static game and also help identify good strategies valid under any setup, including dynamic games. \revisionthree{Because dynamic games, although more realistic, have a significantly larger strategy space, we use reinforcement learning to obtain efficient strategies that lead to equilibrium outcomes.} We show how theoretical findings can be used to train smart agents to play the game and validate our approach in practical settings.
\end{abstract}




\end{frontmatter}



\section{Introduction}
Jamming problems are a typical application of game theory to wireless communications~\cite{perin2021reinforcement}. 
The quintessential jamming scenario \cite{altman2007jamming} involves a zero-sum game played by two agents, i.e., a legitimate node that tries to maximize a utility value related to the efficiency of a transmission and instead a jammer acting as a minimizer \cite{commander2007wireless}.
The utility can be, for example, the available transmission rate over the channel, or, equivalently, the signal-to-noise plus jamming ratio (SNJR) in the case of additive white Gaussian noise (AWGN) channels following Shannon's capacity formula \cite{garnaev2022anti}.

More precisely, we consider a receiver R, located in the proximity of a transmitting access point (AP), the latter being stationary and non-strategic, i.e., not a player in the game. The objective of R is to receive communication from the AP with a high data rate, to make efficient use of the wireless channel. A jammer J is also present in the area, with the intention of disrupting communication, so that R obtains the lowest possible communication rate. We assume that, to pursue this goal, J raises the noise floor of the ongoing communication, thereby causing jamming in the form of additional noise to R's reception. This is a standard adversarial interaction that, from a \ mbox {game-theoretic} point of view, can be framed as a two-player zero-sum game~\cite{dasilva2011game}. Many jamming problems agree on this premise, yet they usually differ in what actions are available to the players, which ultimately makes up for their strategic choices. 

To our knowledge, \revisionthree{most adversarial jamming approaches focus on resource allocation such as channel selection or power control~\cite{gouissem2022accelerated}}. In the literature, there are variations on the technical premises of the communication scenario. For example, J may possess more advanced jamming capabilities to disrupt R's communications, such as eavesdropping or spoofing~\cite{manshaei2013game}. In addition, a reverse situation occurs in the case of \emph{friendly} jamming, where J is a legitimate network agent that wants to disrupt R's transmission, which, in turn, has the malicious intent of stealing data or communicating in a forbidden area~\cite{cai2018joint}.

The aspect of node mobility is not entirely neglected, but overall is rarely addressed in the related literature on game-theoretic jamming. Compared to the case where the strategic aspects relate to power control, assuming that the nodes change position complicates the analysis, as many system parameters are affected, and this increased complexity may explain why this problem is \revisionthree{rarely} addressed. Moreover, another complication to the game-theoretic approach is that the movement of the nodes over time would require \revisionthree{addressing} the strategic interaction as a \emph{dynamic} game.
In the literature, most approaches are static (in the sense of \revisionthree{a} one-shot decision) \cite{chiariotti2020underwater}, or they focus on a two-step dynamic selection in the form of a Stackelberg game \cite{misra2020m}, which still fails to capture a dynamic evolution over multiple time instants and the learning phase of the strategy. 

In this paper, we focus on a full-fledged dynamic game theoretic analysis of mobility strategies in the interactions between R and J. While a straightforward extensive form of the game would rapidly explode, and as such the number of involved strategies becomes prohibitively \revisionthree{large for closed-form analysis}, we can keep the size of the problem contained by introducing some limitation on the mobility of the nodes and the propagation model. We do this while still maintaining all the characteristics that are relevant from a high-level perspective, such as the ability to occupy different positions and the utility being only dependent on the mutual positioning of transmitter, receiver, and jammer, in turn reflecting a diminishing scaling law of the signal quality with distance.

In particular, for the sake of tractability, we focus on nodes moving along a linear geometry; as discussed in the following, this still takes into account all the relevant geometric aspects for the computation of the SNJR \cite{bout2020energy}. In fact, R wants to be as close as possible to the AP while at the same time escaping J, which can instead be considered to chase R \cite{perin2021reinforcement}. 
In line with some other relatively \revisionthree{recent game-theoretic investigations}~\cite {arjoune2020novel}, we also pursue another detour from traditional studies, which allows tackling a harder computational scenario, i.e., the introduction of reinforcement learning (RL) as a tool to find equilibria. We remark that, in light of our usage of RL techniques, the extension to more sophisticated two- or three-dimensional geometries would not make a great conceptual difference from our simpler linear topology, which allows for a more immediate understanding of the analytical findings. 

Thus, the present paper offers the following contributions.
\revision{
\begin{itemize}
    \item \textbf{Analytical solution of the static version of the zero-sum game played by J and R as moving agents.} Despite its apparent conceptual simplicity, it allows the derivation of non-trivial results that serve as the basis of more complex strategic setups with dynamic gameplay. \revisiontwo{A notable result is that the best position for J is close to the AP, to prevent R from visiting regions scarcely affected by path loss.}
    \item \textbf{Introduction of three different dynamic games.} They result from the strategic movements of the involved nodes in more elaborate contexts and, to solve them, two RL methodologies are adopted, based on a classical tabular approach and deep reinforcement learning (DRL)~\cite{sutton2018reinforcement}, respectively. \revisionthree{The three games are described below.}
    \begin{description}
        \item[Game $\mathcal G_1$:] Nodes move by changing their location \emph{sequentially} with \emph{perfect and complete information} about the locations of the other, and individual rewards are cumulated after each player's move.
        \item[Game $\mathcal G_2$:] We iterate \emph{simultaneous} changes of position by both players, making it a game of \emph{imperfect but complete information}. 
        \item[Game $\mathcal G_3$:] In the third and last dynamic game, each player has \emph{imperfect and incomplete information} about the opponent's position. It \revisionthree{only} observes the propagation environment and, therefore, the perceived channel capacity changes as a consequence of the other player's location.
    \end{description}
    \item \textbf{Numerical results.} The simulation campaign conducted for the three dynamic games confirms a good match with the theoretical principles extended from the static game analysis and hints at possible further investigations. \revisiontwo{Specifically, our findings show that complete information is generally beneficial to J, as R's payoff is maximized in game $\mathcal G_3$. When J is unaware of R's position, it tends to remain close to the AP, \revisionthree{consistent} with the intuition developed for the static game.}
\end{itemize}
}
The proposed general setup allows additional aspects to be framed in the model, such as imperfect information about the role of players~\cite{garnaev2020jamming}, their channel conditions~\cite{sagduyu2011jamming}, or their physical location~\cite{scalabrin2015zero}, which can all be included to lead to different Bayesian games. As will be argued, a Bayesian analysis is out of the scope of this paper but can be seen as a further extension of our findings.

For dynamic games, the size of the strategy space and the conceptual difficulty of deriving them in full detail make \revisionthree{a closed-form characterization prohibitive}. However, we are able to identify general principles of strictly dominant strategies, which are rigorously proven and may serve as a guideline for intelligent agents.
Moreover, our adversarial RL approach \revisionthree{can} identify Nash equilibrium (NE) strategies through (deep and non) Q-learning with $\epsilon$-greedy 
exploration-exploitation~\cite{uther1997adversarial}. 

The remainder of this paper is organized as follows. Section~\ref{sec:rel} reviews related work. In Section~\ref{sec:mod} we present the scenario and the mobility model that the nodes use to play. Section~\ref{sec:gam} details the game-theoretic analysis for the static case, which is further extended in Section \ref{sec:dyn} to three different dynamic games and their solution via RL. Section~\ref{sec:res} presents the performance evaluation, and Section~\ref{sec:conclusion} draws the conclusions \revisiontwo{and paves the way for future extensions involving distributed and physics-informed RL.}

\section{Related work}
\label{sec:rel}

Among the many seminal investigations that address jamming through game theory \cite{commander2007wireless}, one of the first systematic game-theoretic studies on the impact of mobility on jamming in wireless communication can be found in \cite{bhattacharya2013jamming}. This paper explicitly addresses mobility as (part of) the strategic choices of the players \revisionthree{in} the context of unmanned aerial vehicles (UAVs), which has some points of contact with our investigation, specifically the idea of modeling the jamming avoidance as a pursuit-evasion \revisionthree{game}.
However, the paper formalizes jamming as a complete-information differential game with continuous-time controls. This abstracts from the information asymmetry that can characterize the gameplay between a jammer and a transmitter, which is something that we try to address and is especially relevant in a terrestrial context, where the position of the nodes may not be fully disclosed. Moreover, \revisionthree{unlike} this fine-grained controllability, we adopt a discrete decision-making approach, which is more robust to strategic uncertainty. 

Other papers actually explore physical layer security for UAVs with similar approaches. For example, a two-layer optimization problem for UAV security, including a non-convex optimization problem in the lower layer and a game theory problem in the upper layer, is formulated in \cite{zhang2025joint}. This paper explicitly addresses mobility as a key component of UAV strategy, but the focus is on counteracting eavesdropping, and network terminals are the ones enacting jamming in a coalition fashion.

Similarly, a UAV scenario is also analyzed in \cite{gao2019anti}, where strategies to counteract jamming are studied, specifically focusing \revisionthree{a Markov decision process solved with deep recurrent Q-networks}. 
\revisionthree{In \cite{qin2025multi}, the authors discuss how multi-agent reinforcement learning can be adopted to find the equilibria in a dynamic game with the objective of preventing jamming, but the focus is again on UAV and non-terrestrial networks.}

Aside from the context of UAVs, where the node positions can be more easily discovered or inferred, there are relatively few studies on mobility for other network scenarios and, in particular, for terrestrial networks.
The choice of a different position to avoid jamming is also a key component in underwater networks, which is the focus of \cite{chiariotti2020underwater}. However, in this contribution, the game is static, as the choices of the players are never changed afterwards, and in the best case, made part of a Bayesian type, without any sort of dynamic update. Another underwater scenario with node mobility used to avoid jamming is studied in \cite{xiao2018anti}, but the approach is not strategic (game-theoretic).

For more general wireless communications, a recent survey \cite{jia2025game} on the use of game theory and reinforcement learning in physical layer security lists many contributions to jamming or similar attacks, but very few among them discuss mobile nodes (either jammer or intended receiver), and none react strategically to each other. 
Exceptions are represented by the approaches of \cite{he2014dynamic} and~\cite{targue2010improving}, based on controlled mobility. However, the work in~\cite{he2014dynamic} does not refer to physical movements and \revisionthree{their} impact on the wireless channel gain but rather focuses on an abstract multi-flow commodity problem. Paper~\cite{targue2010improving} instead modifies the network topology through mobility to obtain better throughput in the presence of jamming attacks.

This lack of game-theoretic characterization for moving transmitters and jammers is a gap that is important to fill since next-generation wireless systems strongly support terminal mobility, not to mention that the overall performance of the wireless channel capacity is heavily affected by movements of users~\cite{aguero2018architectures}. Moreover, the typical reaction of wireless users experiencing poor channel quality is to move away, thus changing position~\cite{xu2007channel}, which still applies in the case of jamming. If the jamming nodes are not strategic and the receiving nodes are allowed to change positions, they will try to move away from the areas with the strongest jamming, as investigated in~\cite{misra2020m}. 

Also, while there exist studies revolving around a specific technology, there is no baseline investigation that, even for a simple radio scenario, considers the evolution of the game on a longer dynamic scale.
For example, in mmWave scenarios, moving around can be seen as a countermeasure to blockage \cite{hussain2020mobility} in addition to the most frequently encountered solution \revisionthree{through} beamforming \cite{zhu2019mitigating}. Still, this rarely considers a strategic perspective, i.e., channel impairments are just random, e.g., based on the stochastic geometry of the obstacles \cite{abdelnabi2020outage}.

\revision{Overall, compared with existing approaches, our framework provides a unified game-theoretic treatment of mobile jamming that jointly captures dynamic mobility, strategic interaction, and different levels of information availability between the players. Unlike prior works that rely on static formulations, continuous-time perfect-information models, or technology-specific assumptions, our approach supports sequential and simultaneous decision-making under both complete and incomplete information, making it better suited to realistic terrestrial wireless scenarios. In addition, the integration of reinforcement learning enables the analysis of dynamic games with large strategy spaces that are analytically intractable, while still preserving interpretable game-theoretic insights.}


\section{Physical Environment and Game Scenarios}
\label{sec:mod}

We consider a wireless receiver R interested in receiving transmissions from an AP\footnote{While we will refer to the transmitter as the ``AP'' throughout the paper for simplicity, we note that our analysis \revisionthree{is not limited} to Wi-Fi, but can be applied to any generic transmitter, e.g., roadside units (RSUs) in the context of \revisionthree{vehicle-to-infrastructure} (V2I) communications.}, while a jammer J plans to disrupt this communication by causing additional noise to R's reception. 
We consider both R and J as mobile nodes that strategically choose their positions, as rational players in a game. 
For a precise formulation of the game, we start by describing the environment in which R and J are placed and move. Subsequently, we consider the wireless channel representation, in particular focusing on the key points for our game-theoretic analysis. 
Indeed, we remark that the required input for a game-theoretic formalization is the availability of the main monotonic trends, such as certain positions being more favorable than others, rather than precisely quantifying the propagation model for any single coordinate. In fact, game theory interprets the comparison of possible outcomes in an \emph{ordinal} \revisionthree{rather than cardinal sense}, which is also expressed by using the NE, i.e., a joint strategic choice from which no player wants to deviate, as a solution concept~\cite{tadelis}.
To introduce rationality in wireless nodes playing a jamming game, it is not required to assume that they know the channel conditions of every position in the scenario, which can be seen as unrealistic; instead, they can just have a good estimate of how the channel gain and interference change as they move \cite{garnaev2022anti}.
We conclude this section by providing examples of technical scenarios in which our analysis would be realistic. This is not meant to limit the scope of our investigation to some technologies and environments, but to prove that our analysis is indeed related to concrete instances of network interactions.

\subsection{Environment Layout}

We focus on a terrestrial scenario, implying that R and J stay on the ground. An extension to drones or devices capable of moving along a vertical coordinate can be envisioned in future investigations, but it would not be conceptually different. Even though they are moving on a surface, the positions of R and J can be represented in polar coordinates $(\rho, \vartheta)$ with the origin at the location of the AP \cite{abdelnabi2020outage}. It is immediate that, while the distance from the AP $\rho$ concretely influences the SNJR perceived by the devices, the angle $\vartheta$ only implies a rotational symmetry of the positions.
Motivated by this reasoning, in this paper we limit the investigations to a linear geometry, which means that all nodes can occupy positions on a positive coordinate $\rho$, taken as the only parameter of interest.
This model, which would be perfectly appropriate for nodes placed along a road \cite{hussain2020mobility}, still allows a description that contains all relevant aspects of our analysis, namely the ability of players to change position. Although using a single coordinate offers the advantage of \revisionthree{notational} simplicity, we will see that the resulting setup already attains a complexity that makes it challenging enough.
Also, our findings can be expanded to more general geometries (e.g., with more directions available). Such an extension can, for example, explore the choice of an angle $\vartheta$ along the value of $\rho$ for both players, but, in general, the analysis would not be affected, especially if the positions available to the nodes are just evaluated from a quantitative perspective of their physical propagation conditions.

\begin{figure}
    \centering
    \includegraphics[width=\columnwidth]{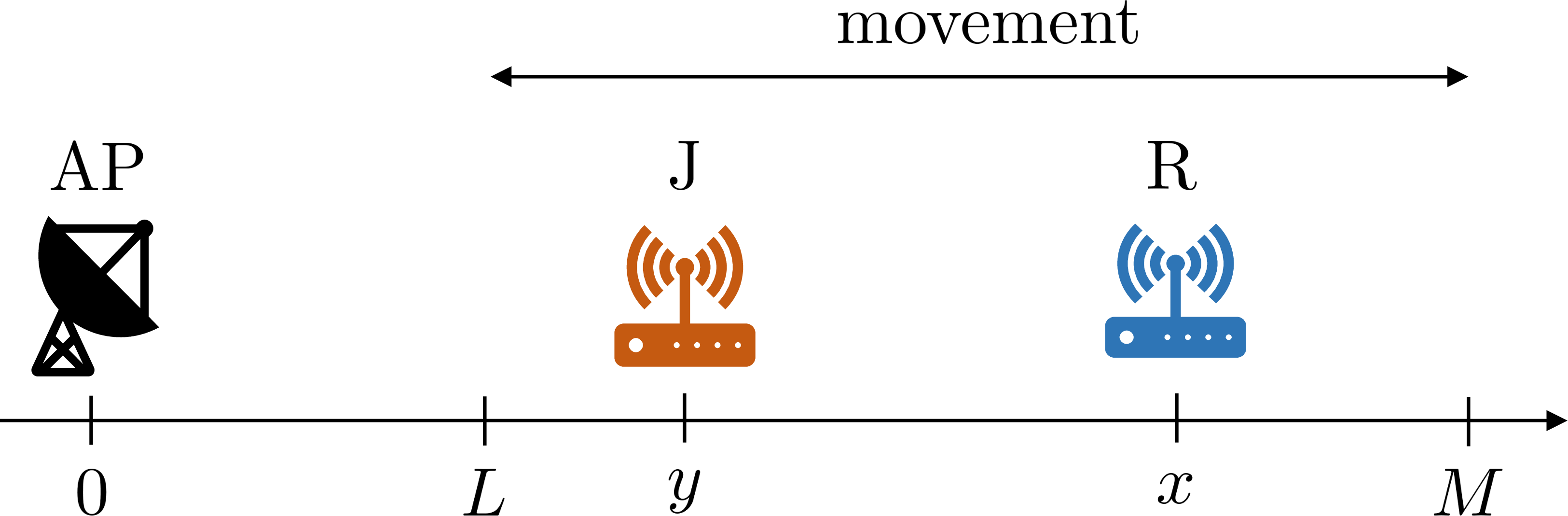}
    \caption{Graphical representation of the considered game, with the access point located at the origin. Players J and R can move between $L$ and $M$.}
    \label{fig:game}
    \vspace{-0.3cm}
\end{figure}

Thus, we consider a half-line starting from an origin $\rho=0$, where three devices are placed. The first is the stationary AP that acts as a transmitter and is not controlled by any player. The AP is placed at the origin, that is, position $\rho=0$, which is in turn forbidden for the other nodes to take. The second and third nodes in the geometric placements are a receiver R and a jammer J, each of them controlled by rational agents that determine positions $\rho_{\rm R} = x$ and $\rho_{\rm J} = y$, for R and J, respectively.
To limit the region where R and J can be placed, we take both $x$ and $y$ as values in the positive coordinate between a minimum value $L$ and a maximum value $M$, with $0 < L < M$. Thus, both R and J can be placed within the segment $[L,M]$ and never too close to the transmitter. Fig.~\ref{fig:game} depicts the scenario. This is only a possible choice for the region where R and J can be located; for example, the region available to the jammer could be further restricted. Nevertheless, all of the conclusions drawn in the following can be promptly extended to other scenarios where the ranges of positions available to the players are different \revisionthree{from} each other.

For propagation, we assume a narrowband AWGN wireless channel between AP and R, with bandwidth $B$, on which J can cause additional noise at the receiver's end. Thus, R experiences a data rate quantified through Shannon's capacity as $C=B \log_2 (1+ \Gamma)$ where $\Gamma$ is the SNJR computed~as
\begin{equation}
\Gamma = \frac{ g_{\rm R} P_{\rm tx}}{\nu_0 + g_{\rm J} P_{\rm J}}
\label{eq:snr}
\end{equation}
with terms $P_{\rm tx}$ and $P_{\rm J}$ being the powers transmitted by the AP and the jammer, respectively, while $g_{\rm R}$ and $g_{\rm J}$ are gain terms between the AP and R, and the jammer and R, respectively, and $\nu_0$ is a noise term.

We apply this physical setup to different games where the set of players always consists of R and J, treated as having contrasting objectives, which is captured by all the games being \emph{zero-sum}~\cite{dasilva2011game}: the \emph{value} of the game is defined as the channel capacity, so that player R has utility $u_{\rm R} = C$, while J has utility $u_{\rm J} = -u_{\rm R} = -C$. This means that R strives for a position pair $(x,y)$ that maximizes the channel capacity $C$, while J wants to minimize it. \revisionthree{Under this formulation}, we can simplify the equations related to the telecommunication scenario by observing that, in game theory, utility functions \revisionthree{represent users' preferences, having no} specific physical meaning in themselves. Thus, \revisionthree{any strictly monotonic transformation} would leave the preferences unaltered since they still respect the principle that the higher the utility, the more preferred the outcome. Especially, if the utility transformation is a linear rescaling, the set of mixed NEs found is the same in the transformed utilities \cite{tadelis}.

In the following, we first obtain two general results valid with and without noise (Theorems~\ref{theo:1} and~\ref{theo:NEgen}). Then, we set the noise term $\nu_0 = 0$, which is reasonable as the jamming effect is expected to be preponderant, and ignore the bandwidth $B$ and the differences between the transmitted powers $P_{\rm tx}$ and $P_{\rm J}$, which only cause a proportional rescaling of the capacity. Under these assumptions, we are left with a value that is a logarithmic function of $\log_2 (1+\Gamma)$; therefore, we can exploit the equivalence $\log (1+a) \sim a$ for small $a$. After another proportional rescaling due to the change in base for the logarithms, we find that the value of the game just depends on the SNJR, boiling down to the ratio between $g_{\rm R}$ and $g_{\rm J}$, and ultimately to the users' positions $x$ and $y$ as
\begin{equation}
\text{value} = u_{\rm R} = -u_{\rm J} = \frac{ |x-y|^\alpha }{x^\alpha}.
\label{eq:valfin}
\end{equation}
This is not surprising and, in fact, the same framework can be used for other metrics of interest beyond capacity $C$. For example, in~\cite{garnaev2022anti}, a similar approach is taken to support a latency-based analysis, since, when the value of the SNJR is small, it is inversely proportional to the latency. With these assumptions, it is possible to obtain the closed-form results of Theorem~\ref{theo:NE}. We also discuss what happens when the noise is not neglected in Remark~\ref{rem:noise} and provide more insights with the results of the dynamic games in Section~\ref{sec:res-net-params}.

Some remarks are now in order. First, more complex propagation models can be employed, but this would only make the analysis more complicated, while the meaning of the communication performance would still be valid on average. Moreover, the dependence on $\alpha$ is also marginal, since it will not alter the ordinal meaning of the utilities. Other results \cite{chiariotti2020underwater,scalabrin2015zero} in the field of games with variable positions of users hint at this principle by numerically confirming it. For the sake of choosing a value, in the following, we will most often refer to the case $\alpha=2$, which corresponds to propagation in free space. However, it will be clear that the results obtained are more general, and we emphasize that the closed-form results in the analytical formulation actually do not depend on the numerical value of~$\alpha$.

Even in the case that a more detailed channel model is employed, such an extension would only complicate the problem possibly to the point of becoming computationally intractable, without altering the fundamental result, i.e., that the relative positions of the jammer and receiver remain the key determinant of system performance in the context evaluated here. Thus, model realism is intentionally not the focus nor the primary contribution of this paper. However, in Section~\ref{sec:802.11p} we provide evidence that our simple model captures the average behavior of the wireless channel. Furthermore, it is not our goal to propose any novel RL framework but rather to demonstrate via RL that the analytical solutions derived for the static setting still apply to the more complex dynamic setting (see Section~\ref{sec:res}), which is otherwise too complex to be treated analytically.

\subsection{Application to IEEE 802.11p}
\label{sec:802.11p}
A \revisionthree{relevant application scenario} of the proposed framework is related to vehicular networks. The IEEE 802.11p standard, working in the $[5.850,5.925]$ GHz subband~\cite{abdelgader2014physical}, can be considered a good example of a communication scenario where nodes are inherently mobile and are required to exchange dedicated short-range communication (DSRC) messages for traffic assistance. In this context, cooperative awareness messages need to be exchanged to gain information about the traffic context and allow vehicle support, such as avoiding congestion and reducing accidents~\cite{mangel2011real}. In the context of autonomous vehicles, this is envisioned to enable safe navigation throughout an urban environment. However, it would be extremely easy for a malicious jammer to cause interference to the related communication exchanges even without particularly sophisticated hardware. For example, IEEE 802.11ac can use extended Wi-Fi channelization at $5$ GHz, spanning the \revision{industrial, scientific, and medical (ISM)} band, which can cause interference to IEEE 802.11p~\cite{khan2017can}. Thus, an intentional jammer can jeopardize the communications of the legitimate receiver by simply raising the noise floor and impeding reception using commercial equipment, which is precisely our context. 

\begin{figure}
    \centering
    \includegraphics[width=\columnwidth]{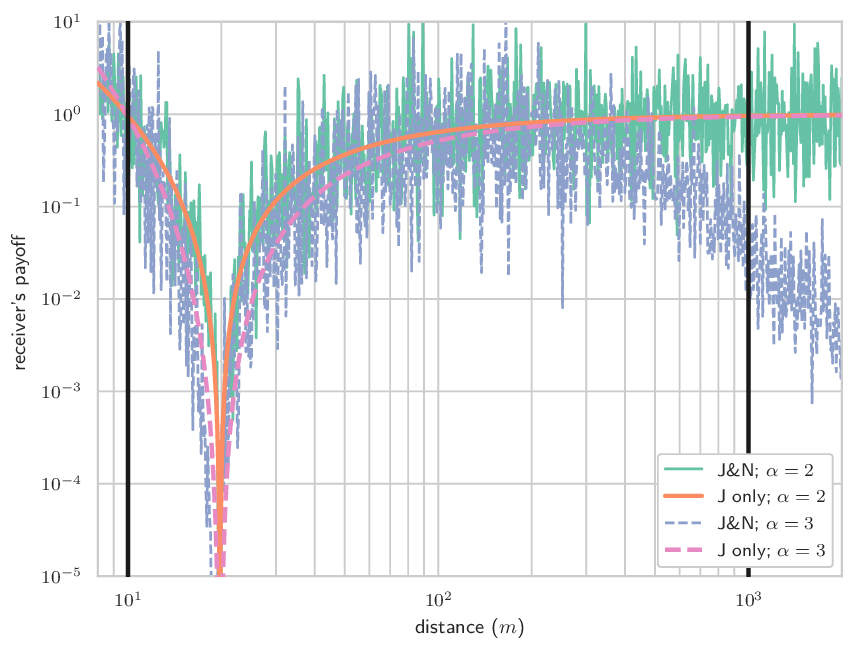}
    \caption{Model used and its relation to a realistic scenario. The black vertical lines denote the values of $L=10$ m and $M=1000$ m.}
    \label{fig:realistic_payoff}
    \vspace{-0.3cm}
\end{figure}

According to the usual dimensioning of these networks in practical on-field evaluations (see, for example,~\cite{mangel2011real} \revisionthree{or} \cite{qin2014performance}) we can argue that our analysis neglecting the presence of noise fits well a vehicular scenario where the distances that the mobile nodes can span are around $1000$~m if the line of sight (LoS) is available, and a fifth of that in a non-LoS scenario. Thus, we can think of applying our analysis with parameters such as $L=10$, $M=1000$ meters, and using exponents $\alpha=2$ and $\alpha=3$ for LoS and non-LoS, respectively. In Fig.~\ref{fig:realistic_payoff} we show the model obtained from the approximation of \revision{Eq.~\eqref{eq:valfin}} when J \revision{plays its best response $y^\star\simeq 19.8$~m} (see Section~\ref{sec:gam}, \revision{Theorem~\ref{theo:NE}}). The realistic vehicular scenario is obtained for IEEE 802.11p with the presence of log-distance correlated shadowing ($\sigma^2_{\rm{LoS}}=3$, $\sigma^2_{\rm{nLoS}}=4$) and a noise floor $\nu_0=-101$~dBm (obtained using $B=20$~MHz, \revisionthree{specified by the standard}, and a normalized noise floor $\nu_{\rm B}=-174$~dBm/Hz at \revisionthree{ambient} temperature. The transmission powers of the AP and the jammer are $P_{\rm tx}=P_J=23$~dBm. As can be seen, for $\alpha=2$ the proposed model can capture the average behavior of the environment throughout the considered range. In non-LoS conditions, with a higher path loss exponent, the model is reasonably valid up to a distance of about $200$~m, in correspondence with the maximum present considering the action of the noise. After this value, the receiver's payoff starts to decrease because the noise floor has a comparable impact \revisionthree{to} the interference caused by the jammer (which has reduced due to the distance). In the latter case, one should consider smaller values of $M$ to obtain consistent results.

These considerations are not limited to vehicular scenarios, but also hold, albeit with slightly different models in terms of timescale, speed, LoS vs.\ non-LoS propagation, to a wide array of networks exploiting \revision{radio-frequency (RF)} communication \cite{bout2020energy}, mmWave \cite{zhang2021secure}, satellite-aided positioning \cite{medina2019gnss}, or underwater acoustic~\cite{chiariotti2020underwater}.
Moreover, they still apply, to some extent, if the set of available positions is limited~\cite{zhu2019mitigating} and/or the legitimate network agents can approach the jammer to stop it~\cite{perin2021reinforcement}.

\section{Static Game Analysis}
\label{sec:gam}
First of all, we study a static version of the game, where R and J just choose, independently and unbeknownst to each other, a location in $[L, M]$ and stay there forever. This game does not involve any mobility, but the free choice of locations sheds light on how R and J are supposed to behave in a mobile scenario. The framework of a static game is generally tractable, as it can be categorized as a \emph{zero-sum game on the unit square with polynomial utilities}~\cite{glicksberg20169}.
\revision{It follows that analytical conclusions can be drawn if we formalize a static game $\mathcal{G}_0$ as follows: 
\begin{enumerate}
    \item the set of players is R and J;
    \item their action set consists of the interval $[L,M]$, in which they both make a unique simultaneous choice;
    \item for the results in Theorems~\ref{theo:1}--\ref{theo:NEgen}, we assume the final utility is indifferently the channel capacity, the spectral efficiency, or the SNR of Eq.~\eqref{eq:snr};
    \item for the result of Theorem~\ref{theo:NE}, where a closed-form NE is retrieved, we assume the final utility is the value of the game as per~\eqref{eq:valfin}, with general path loss exponent $\alpha \geq 1$.
\end{enumerate}}
We obtain the following results.

\begin{theorem}
\label{theo:1}
In game $\mathcal{G}_0$, the best response of the jammer to the belief that player R is choosing a pure strategy is to choose the same strategy as R. Moreover, R cannot play a pure strategy at an NE. Hence, any NE must be mixed from R's side.
\end{theorem}
\begin{proof}
See \ref{app:1}
\end{proof}

\revision{
\begin{theorem} 
\label{theo:thresh}
There exists a value $\tilde{M} \in (L, M]$ such that it is not rational for R to play a strategy $x > \tilde{M}$. 
\end{theorem}}
\revision{\begin{proof} 
See \ref{app:1b}
\end{proof}}
\revision{
This theorem states that the effective strategy space of R is $[L,\tilde{M}]$, because R would not place itself beyond $\tilde{M}$.
Note that it can be that $\tilde{M} = M$, which is what happens in the absence of noise, or if the physical limit to how far R can go without violating the theorem lies beyond M. However, for large $\nu_0$ and large $M$, the set of positions occupied by a rational R is narrower than $[L,M]$ as $\tilde{M} < M$.}

\begin{theorem}
\label{theo:NEgen}
$\mathcal{G}_0$ admits a single NE, where
\revision{
\begin{enumerate}[(i)]
    \item J's best response is to play $y = \revision{y^\star}$ with $L < \revision{y^\star} <M$ with probability $1$;
    \item R's best response is to choose a mixed strategy whose support only contains $L$ and $W\le\tilde{M}$.
\end{enumerate}
}
\end{theorem}
\begin{proof}
See \ref{app:2}
\end{proof}

\revision{\begin{remark}
\label{rem:noise}
If $\tilde{M} <M$, which requires the presence of noise, the single NE of $\mathcal{G}_0$  corresponds to J choosing $y=y^\star_{\rm noise} < \revision{y^\star}$ and R playing mixed strategy $x=pL+(1{-}p)W$ for some $0{<}p{<}1$. 
This implies that Theorem \ref{theo:NE} still holds, but $M$ is replaced by a smaller $W \leq \tilde{M}$ due to the effect of noise. To compute $W$ precisely, one has to resort to numerical methods by leveraging the indifference criterion. For example, if the utility function is Shannon's capacity, i.e., $u_R = B\log_2(1+\Gamma)$, the indifference criterion $u_{\rm R}\big(L,y^\star\big) = u_{\rm R}\big(W,y^\star\big)$ yields
\begin{equation}
    \begin{aligned}
        \label{eq:indifference}
        B\log_2\left(1+\frac{P_{\rm tx} g_R(L, y^\star)}{\nu_0+P_J g_J(L,y^\star)}\right) &= B\log_2\left(1+\frac{P_{\rm tx} g_R(W,y^\star)}{\nu_0+P_J g_J(W, y^\star)}\right) \\
        \iff \frac{P_{\rm tx} g_R(L, y^\star)}{\nu_0+P_J g_J(L,y^\star)} &= \frac{P_{\rm tx} g_R(W, y^\star)}{\nu_0+P_J g_J(W,y^\star)}\\
        \iff L^\alpha \left[\nu_0 + P_J(W-y^\star)^\alpha\right] &= W^\alpha \left[\nu_0+P_J(y^\star-L)^\alpha\right]
    \end{aligned}
\end{equation}
The practical characterization of this remark is that the presence of noise makes positions further from the AP less attractive to R. Reacting to this, J will choose a position closer to the AP, i.e., $y=y^\star_{\rm noise} < \revision{y^\star}$. The best response of R, according to Theorem~\ref{theo:NEgen}, must be a mixed strategy playing opposite coordinates, but the support contains $L$ and a smaller value $W$ instead of $M$. To visualize this, refer to Fig.~\ref{fig:realistic_payoff}: in the case with noise and $\alpha=2$ (solid green line), we have $\tilde{M}=M$, and then $W=M$; on the other hand, when $\alpha=3$ (dashed blue line), the best payoff for R is observed before $M$ ($W < M$).
\end{remark}}

\begin{theorem}
\label{theo:NE}
When the noise term $\nu_0$ tends to $0$ (in which case $\tilde{M}=M$), the single NE of $\mathcal{G}_0$ is
\revision{
\begin{enumerate}[(i)]
    \item J choosing the best response $y = \revision{y^\star}$ with
    \begin{equation}
        \label{eq:j-optimal-position}
        y^\star = \frac{2LM}{L+M};
    \end{equation}
    \item R playing the best response $x=x^\star$, with
    \begin{equation}
        \label{eq:r-optimal-position}
        x^\star =
        \begin{cases}
            L & \text{with prob. } p^\star = \frac{L}{L+M},\\
            M & \text{with prob. } 1-p^\star = \frac{M}{L+M}.
        \end{cases}
    \end{equation}
\end{enumerate}
The value of the game at the NE is 
\begin{equation}
    \label{eq:game-g0-value}
    u_{\rm R}^\star = \left(\frac{M-L}{M+L}\right)^\alpha.
\end{equation}
}
\end{theorem}
\begin{proof}
See \ref{app:3}.
\end{proof}

\revision{
\begin{remark}
    We note that the result of Theorem~\ref{theo:NE} is valid under the assumption $g_J P_J \gg \nu_0$ (so that $\nu_0$ can be neglected in the computations) for any value of the other network parameters. Eq.~\eqref{eq:indifference} with $\nu_0=0$ allows us to cancel out also $P_J$.
\end{remark}
\begin{remark}
    The network parameters $B$ and $P_{\rm tx}$ only cause the value of the game $u_{\rm R}^\star$ to be rescaled, but the NE $(x^\star,y^\star)$ would remain the same regardless of the specific choice.
\end{remark}
}
\begin{figure}
\centering
\includegraphics[width=\columnwidth]{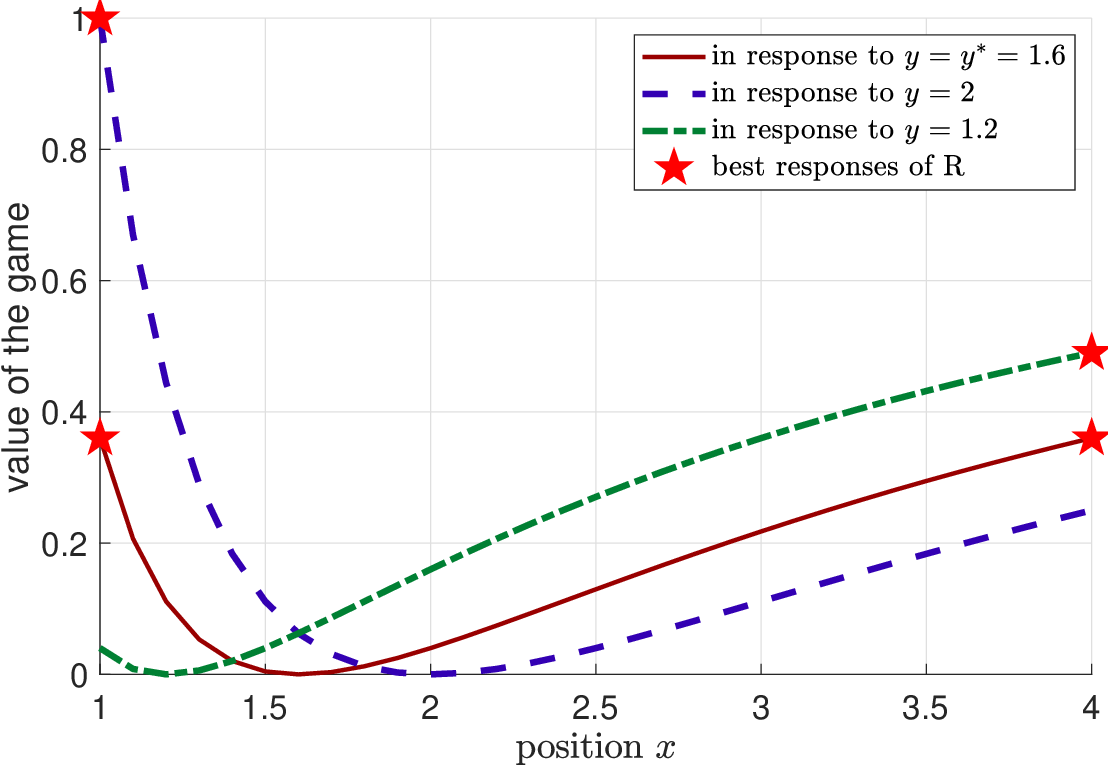}
    \caption{Value of the game $u_{\rm R}$ versus the position $x$ chosen by the receiver R, for various positions $y$ of the jammer J.}
    \label{fig:223}
    \vspace{-0.3cm}
\end{figure}

Hence, it turns out that the best strategy for R is to choose a mixed strategy with a combination of the extreme locations $L$ and $M$, to which the best response from J is to place itself in a specific intermediate position in the interval $[L,M]$.
In the context of a static game, this mathematical result is justified by the adversarial setup. Since the receiver tries to be as far from the jammer as possible, the best positions are the extreme locations $x=L$ and $x=M$. However, since choosing either of them with certainty is clearly disadvantageous, R must \revisionthree{instead} randomize between them, and as a result, J chooses an intermediate position to cover both situations. 

A graphical visualization of Theorem \ref{theo:NE} is shown in Fig.\ \ref{fig:223}, where we plot the value of the game depending on the choices of the players.
It is shown that J's choices of $y<\revision{y^\star}$ and $y>\revision{y^\star}$ imply R's best response to being $x=M$ and $x=L$, respectively, while $y=\revision{y^\star}$ sets R's indifference between them and also causes the overall lowest value, thus making it J's preferred choice.

As a further result, we note that a similar situation holds even when the choices of the players are not simultaneous. This corresponds to what in game theory is known as a Stackelberg game \cite{lmater2018smart}, i.e., one player moves first, making the move public to the second player that acts with full knowledge of it.

If we consider a Stackelberg version of $\mathcal{G}_0$, i.e., a version with \revisionthree{sequentially revealed} choices, the first player to move is called \emph{leader}, and the second is the \emph{follower}. Then, the following theorem holds.

\begin{theorem}
\label{theo:Stack}
Consider a Stackelberg version of $\mathcal{G}_0$ with $\nu_0 \rightarrow 0$. The following results hold.
\begin{itemize}
\item[i)] When J is the leader, J's best response is to choose $y=\revision{y^\star}$, while R's best response is any linear combination $x=pL+(1-p)M$, with $0\le p \le1$. As a consequence, the NE described in Theorem \ref{theo:NE} is also a Stackelberg equilibrium for game $\mathcal{G}_0$.
\item[ii)] When R is the leader, the resulting Stackelberg equilibrium is also the same as the NE only if R is allowed to choose a mixed strategy. In pure strategies, the maximin of player R (and the minimax of player J) is $0$.
\end{itemize}
\end{theorem}

\begin{proof}
See \ref{app:4}.
\end{proof}

The analysis of the Stackelberg equilibrium reveals what happens if the moves are not simultaneous but rather players somehow reveal their decisions after making them. It turns out that the scenario is advantageous for the jammer, as the same NE is achieved even when its position is revealed. In other words, even if R discovers that there is a jammer located at $\revision{y^\star}$, it has no effective countermeasures that improve over the NE. Conversely, if R's position is disclosed (as a pure strategy), J obtains a significant advantage because it \revisionthree{achieves} the best possible value (from a minimizing perspective) by choosing that position as well. 

In practical scenarios, there are other advantages for the jammer. For example, J's best choice $\revision{y^\star}$ is robust to variations of $L$ and $M$. 
In addition, choosing a mixed strategy whose support consists of $x=L$ or $x=M$ may not be realistic for R, as it implies that R can choose either side of a long road at will. If R actually chooses one location by moving towards it through subsequent steps, this is more convenient for a strategic jammer. In fact, J could even consider improving on $\revision{y^\star}$ by moving towards R, since (see again Fig.\ \ref{fig:223}), despite it being the best static response, once J knows whether R chose to play $x=L$ or $x=M$, it can move toward the very same position, and R \revisionthree{has no effective escape strategy}. Another consideration, which can be exploited in the analysis of incomplete information scenarios, can be obtained by setting different values of $M$. This shows another strategic advantage of the jammer, as the value of $\revision{y^\star}$ as per Theorem \ref{theo:NE} tends asymptotically to $2L$ as $M \to \infty$. This means that even if R is allowed to move very far, J can minimize the value by always choosing a position between $L$ and $2L$. This reasoning strongly supports the need for further analysis involving dynamic games, which will be covered in the next section. We argue that our findings possibly cast \revisionthree{doubt} on the validity of most of the literature where \mbox{game-theoretic} approaches are only considered in a static context, whereas, as will be shown next, dynamic games develop very differently.

\section{Dynamic Games}
\label{sec:dyn}

We now consider a dynamic setup in which players are allowed to \revisionthree{change their positions over time}. For tractability reasons, we take a similar approach to most investigations in the literature on dynamic games, i.e., we consider a discrete state space~\cite{tadelis}. Thus, we assume that $N$ positions are available to the mobile nodes, within the extremes $\rho=L$ (closest to the AP) and $\rho=M$ (furthest), all equally spaced. The positions are quantized with a step $\Delta = ({M-L})/({N-1}).$ Depending on the value of $N$ and the physical size of $L$ and $M$, we can define a different extent of the mobility pattern. In the following, we will assume that the players are allowed to move up to $S$ steps in both directions from their current position (provided that they do not go beyond the borders). Clearly, the choice of $S$ involves a trade-off, since setting it at any value greater than or equal to $M-L$ implies that the nodes can change their position at will. Also, in~\cite{perin2021reinforcement}, we set $S=1$. To further extend the results, we will also discuss the situation $S=2$ in this paper. 
When we investigate a dynamic setup, we can obtain several different formulations of the game, especially revolving around the order of movement of the players, which can be either sequential or simultaneous, and the information about the opponent's position. Formally, they can be represented as Markov decision processes (MDPs), i.e., discrete-time control processes constituted by the tuple $(\mathcal{S}, \mathcal{A}, \mathcal{P}, \mathcal{R})$, whose elements correspond to the sets of states, actions, transition probabilities, and rewards, respectively. \revisionthree{We use model-free RL to learn state-action values directly from observed transitions and rewards.} Thus, we formalize three dynamic games, whose common features are i) the reward $\mathcal{R}$, given according to the utility function of R and J (Eq.~\eqref{eq:valfin}), and ii) the action set $\mathcal{A}_S$, which depends on the value $S=\{1,2\}$. Specifically,
\begin{equation}
\mathcal{A}_S =
\begin{cases}
    \{\text{stay, move right, move left}\} & \text{if } S=1,\\
    \mathcal{A}_1 \cup \{\text{move right 2, move left 2}\} & \text{otherwise}
\end{cases}.
\end{equation}
\revision{The three games \revisionthree{differ} according to the information set available for the decision-making procedure or for the order of play. \revisionthree{Specifically,}
\begin{description}
    \item[\emph{Game }$\mathcal{G}_1$:] Players move sequentially, i.e., taking turns with alternating moves. They start from a uniformly random position, and the first player to move is also chosen at random. After each move, the player collects its payoff for that round. The players are always informed of the positions of each other, so the state of the game is $\mathcal{S}_1=(x_t,y_t)$, and it can be seen as a common type of the players~\cite{tadelis}. Thus, this is a dynamic game of complete and perfect information.
    \item[\emph{Game }$\mathcal{G}_2$:] Similar to the previous case in that the positions are known to both players and they start from a position chosen at random. However, moves are now simultaneous, which can alternatively be seen as the information set of each player $i$ at time $t$ comprising the positions of both players up to time $t-1$, but only that of player $i$ at time $t$, while the current position of the opponent $-i$ is not distinguishable, i.e., $\mathcal{S}_2=(x_t,y_{t-1})$. After each move, both players collect their payoff for that round. Compared to the previous case, this is a game of complete but imperfect information, since positions are eventually known, but at each turn the move of the adversary is not disclosed immediately.
    \item[\emph{Game }$\mathcal{G}_3$:] The players have no information on the position of the opponent at any past, present, or future time, namely, $\mathcal{S}_3=x_t$. Thus, this is a game of incomplete information~\cite{tadelis}. Even though the order of moves is \revisionthree{irrelevant for the available information, for definiteness}, we assume that they play simultaneously as per $\mathcal{G}_2$. Importantly, while the position of the opponent is not known, the following facts are all common knowledge among the players: (i) there is an opponent; (ii) it must be in a specific position between $L$ and $M$; (iii) it can move one step at a time; (iv) propagation effects are as described in Section \ref{sec:mod}. From the point of view of the following analysis, this scenario is configured as a partially-observable system~\cite{hussain2020mobility}. 
\end{description}
}

The space complexity of these games is too high for a precise closed-form representation. For example, the exact definition of the strategies for each player would involve tracking the entire history of the game, whose representation \revisionthree{grows rapidly with the game horizon}.
Tie-breaking criteria would be required, which would result in a very heavy formalization. Yet, we can characterize some strategy classes such as ``approach'' or ``escape from'' (i.e., moving in the opposite direction of) the other player, with intuitive meaning. Accordingly, we can formalize some general principles, \revisionthree{stated in} Theorems \ref{rem:1} and \ref{rem:2} below.
\begin{theorem}
\label{rem:1}
If R's position is known to J, ``escape'' is a strictly dominated strategy for J.
\end{theorem}
\begin{proof} 
See \ref{app:5}.
\end{proof}
\begin{theorem}
\label{rem:2}
If R's position is unknown to J, the class of strategies where J stays between $y=L$ and $y=2L$ guarantees an upper bound on the value as $u_{\rm R} \leq 1$, regardless of R's actual position and/or the value of $M$.
\end{theorem}
\begin{proof} 
See  \ref{app:6}
\end{proof}

This latter theorem implies that, in the case of unknown positions (and especially if the boundaries of the available positions are not well known to the players), J ought to stay in $[L, 2L]$ (i.e., at relatively low coordinates), since R is expected to move far from $x=L$ and toward $x=M$, as it wants to move away from the jamming. Overall, this can be a useful building block for a practical strategy to play in the case of incomplete information. However, this does not exclude an exploration component for J's moves, to further locally reduce the value.

Beyond these formal findings, for tractability reasons, it may be convenient to take an alternative approach beyond purely analytical computations to assess how the system behaves in dynamic cases. Specifically, we resort to reinforcement learning~\cite{sutton2018reinforcement} to investigate an adversarial scenario that involves players R and J as learning agents. We assume that the two agents learn their policy concurrently, and the three different cases $\mathcal{G}_1$--$\mathcal{G}_3$ are considered. \mbox{Q-learning} with an \mbox{$\epsilon$-greedy} \mbox{exploration-exploitation} policy is used to train the agents' policies online. Specifically, the algorithm relies on the Bellman recursive update of the \mbox{so-called} \mbox{state-action} value function
 \begin{equation}
    \label{eq:q_learning}
    Q(a_t,s_t) = (1{-}\ell)\,Q(a_t,s_t)
    +\ell\Big(r{+}\gamma\max_a Q(a,s_{t+1})\Big)
\end{equation}
where $s \in \mathcal{S}$ and $a\in\mathcal{A}$ are the observed state of the system and the action chosen by the agent, while $t$ refers to the time index, $\ell$ is the learning rate, and $\gamma$ is the discount factor that accounts for the relevance of the future. Therefore, both players keep a table in which they update this function, which corresponds to the expected long-term reward of taking action $a_t$ while in state $s_t$. Note that in addition to the \revisionthree{difference} between simultaneous and sequential moves, the three analyzed cases also differ because of the input state information. In games $\mathcal{G}_1$ and $\mathcal{G}_2$ the state for each agent is the tuple $(x,y)$ (complete information), while the input in game $\mathcal{G}_3$ (incomplete information) is represented by either element of the pair of positions, the one corresponding to the position of the moving player.

Since the general goal of players is to maximize their own payoff, their best policy is to choose the action maximizing the \mbox{$Q$-function} in a certain state, i.e.,
\begin{equation}
    \label{eq:greedy}
    a_t^\star = \argmax Q(a,s_t).
\end{equation}
Motivated by this, one can define the state value function, which corresponds to the average long-term reward obtained following the best policy in a given state, i.e.,
\begin{equation}
    \label{eq:value}
    V(s_t) = \max_a Q(a,s_t).
\end{equation}
Rule~\eqref{eq:greedy} is named \emph{greedy} policy. However, because the agent must also \emph{explore} the environment's responses, it is convenient to add a certain amount of random behavior. Hence, an \mbox{$\epsilon$-greedy} policy uses~\eqref{eq:greedy} to select the action with probability 1-$\epsilon$, and chooses a random action otherwise (prob.~$\epsilon$). Possibly, the value $\epsilon$ is decaying over time, as it is more important to explore the environment at the beginning. The considered task \revisionthree{becomes challenging due to} the presence of an opponent that modifies the response of the environment in a \mbox{non-stationary} way, as it concurrently learns its policy. Although several works, exploiting both traditional \cite{uther1997adversarial} and deep RL \cite{pinto2017robust}, have shown that \mbox{Q-learning} can be outperformed by more complex algorithms in adversarial contexts modeled through Markov games, we chose it for its simplicity. In Section~\ref{sec:res_conv}, we provide evidence that it is good enough to achieve convergence in a \mbox{proof-of-concept} scenario. Besides, we evaluate a more advanced dueling deep Q-network (DDQN)~\cite{wang2016dueling} approach that uses a neural network (NN) to approximate the Q-function as
\begin{equation}
    \label{eq:ddqn_approx}
    Q(a,s_t) = V(s_t) + A(a,s_t),
\end{equation}  
where $A(a,s_t)$ is the \emph{advantage} value of taking action~$a$ in state~$s_t$. While it is known that deep reinforcement learning (DRL) is not \revisionthree{theoretically guaranteed} to converge, empirically good performance has been shown for scenarios like ours, as well as an enhanced capability in handling high-dimensional problems. We, therefore, apply this approach to deal with more complex scenarios where the state-action space is larger.

\section{Numerical Results}
\label{sec:res}

Since the static game $\mathcal{G}_0$ (even in a Stackelberg version) is fully characterized through Theorems \ref{theo:1}--\ref{theo:Stack}, we can limit practical evaluations to dynamic games $\mathcal{G}_1$--$\mathcal{G}_3$ that offer less immediate conclusions.
In the following subsections, we discuss the settings and practical results obtained.\footnote{\revision{The code is released in open source at \url{https://github.com/gioperin/jammingGames}.}}

\subsection{Scenario settings}

We investigate dynamic games $\mathcal{G}_1$--$\mathcal{G}_3$ in a scenario with discrete positions in the range $[L,M] = [10,50]$. Specifically, we take $N=9$ positions in $\{10, 15, \dots, 50\}$ and step $\Delta=5$. We assume that the nodes can move at most $S=2$ positions in either direction. Under these numerical assumptions, there are $N^2=81$ possible states when players act in the joint space (namely, $\mathcal{G}_1$ and $\mathcal{G}_2$), and $N=9$ states for $\mathcal{G}_3$. Throughout the simulations, we set $\alpha=2$ to obtain a consistent model with realistic communications according to what was said in Section~\ref{sec:mod}.

Moreover, there are $5$ available actions for $\rho \in \{ L+2\Delta, \dots, M-2\Delta \}$, i.e., stay still, and move left or right of one or two positions (if $S=2$), unless the position is close to the border values, \revisionthree{subject to the boundary constraints} (i.e., in $\rho \in \{L,M\}$ there are only $3$ moves available, and $4$ for \revisionthree{$\rho \in \{L+\Delta,M-\Delta\}$.} This means that each player has to explore up to $1521$ values in the joint space, and $39$ values instead, in the case $\mathcal{G}_3$ with incomplete information. The simulation environment is run for $1.5 \times 10^6$ iterations, with discount factor $\gamma=0.99$ and learning rates $\{10^{-4}, 5 \times 10^{-2}, 10^{-2}\}$ for games $\mathcal{G}_1$, $\mathcal{G}_2$ and $\mathcal{G}_3$, respectively, in the case of the classic Q-learning approach. A learning rate of $10^{-4}$ is used for the NN, \revisiontwo{and the network is trained with ReLU activation functions and RMSprop optimizer.} \revision{Other relevant parameters are listed in Tab.~\ref{tab:parameters}.} A \revision{custom} $\epsilon$-greedy policy is adopted, decaying the value of $\epsilon$ \revision{with} a hyperbolic cosine function until the minimum value of $\epsilon_{\rm{min}}=0.01$, \revision{according to the formula
\begin{equation}
    \label{eq:cosh-decay}
    \epsilon(t) = \max\left\{\epsilon_{\min}, 1 -\text{sech}\left(\exp\left(-\frac{t-AE}{BE}\right)\right)\right\},
\end{equation}
with $A=1/2$, $B=1/4$, and $E$ being the number of epochs. The rationale behind this choice in place of the more common exponential decay is that the rule~\eqref{eq:cosh-decay} ensures more exploration at the beginning and a steeper transition towards exploitation towards the end of training.} The games are evaluated based on the average reward obtained by the receiver, which is a proxy of the channel capacity according to \eqref{eq:valfin}. Hence, the higher the reward for player R, the higher the data rate.

\begin{table}[tbh]
\centering
\resizebox{0.9\columnwidth}{!}{
\revisiontwo{
\begin{tabular}{@{}lc p{5cm}@{}}
\toprule
\textbf{Parameter} & \textbf{Value}          & \textbf{Description}                             \\ \midrule
$\eta$             & $10^{-4}$               & Learning rate                                    \\
$\gamma$           & $0.99$                  & Discount factor                                  \\
$\epsilon_{\min}$  & $0.01$                  & Min exploration probability                      \\
$B_{\rm size}$     & $128$                   & Batch size                                       \\
$M_{\rm size}$     & $10^4$                  & Replay memory buffer size                        \\
$E_{\rm total}$    & $10^5$                  & Number of episodes                               \\
$E_{\rm replace}$  & $10^3$                  & Number of episodes to replace the target network \\
$H_{\rm layers}$   & $3$                     & Number of hidden layers                          \\
$N_{\rm neurons}$  & $512$                   & Number of neurons                                \\ 
$N_{\rm runs}$     & $10$                    & Number of independent runs                       \\
$\rm{seeds}$       & $\{0,\ldots, 9\}$       & Random seeds for the ind. runs
                    \\
\bottomrule
\end{tabular}}
}
\caption{Summary of the parameters used for the DDQN.}
\label{tab:parameters}
\end{table}

\subsection{Stability convergence}
\label{sec:res_conv}
\begin{figure}
\centering
\resizebox{\columnwidth}{!}{\input{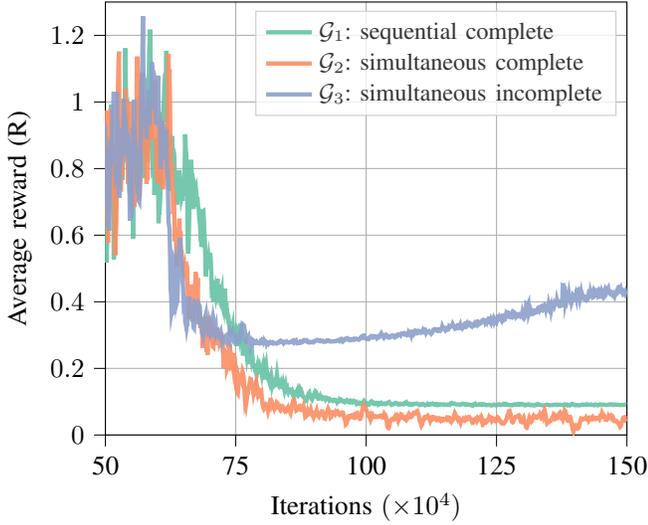}}
    \caption{Receiver's average reward of the tabular \mbox{Q-learning} in the three cases. The result is filtered with a moving average of width $W=5000$ samples.}
    \label{fig:convergence}
\end{figure}
In Fig.~\ref{fig:convergence}, R's instantaneous reward in games $\mathcal{G}_1$--$\mathcal{G}_3$ with $S=1$ is plotted as a function of time (iterations) for the three cases. The plot shows post-processed values after a moving average (MA) filtering, using a window of size \mbox{$W=5000$}. As stated before, the three curves stabilize, which provides evidence of convergence towards a stable solution: this happens around mid-simulation for complete information games ($750 \times 10^3$ iterations). Game $\mathcal{G}_3$ shows an oscillatory behavior as each of the players adapts to the strategy of the opponent. This is due to the nonstationarity of the environment: the players change the distribution of their reward with their moves, unknown to the opponent. Formally, this scenario can be modeled as a partially observable Markov decision process \revisionthree{(POMDP)}. Specifically, if the opponent's position is unknown, J is the first player to find a near-optimal policy, while R learns its best response as a consequence. Therefore, a minimum is observable at around iteration \mbox{$800 \times 10^3$}. Moreover, among the three cases, the one with incomplete information is the most advantageous for R, as J does not know R's position and is therefore prevented from following it closely (see Theorem~\ref{rem:2}). On the other hand, the simultaneous game with complete information is advantageous for J, because it knows R's position and can forecast its moves, always keeping close to it. The sequential game with perfect information places itself in the middle: J observes R's position but can only move reactively, allowing for a slight improvement in terms of payoff for~R.

\subsection{Players' policies in the three dynamic games}

\begin{figure}
    \centering
    \includegraphics[width=\columnwidth]{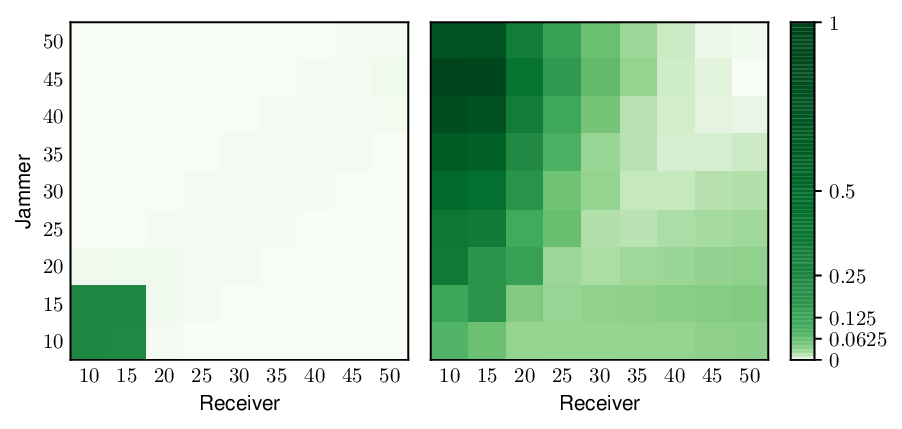}
    \caption{Game $\mathcal{G}_1$, sequential game with perfect and complete information. Joint probability for R and J to find themselves in position $(x,y)$ (left). State value function of R, i.e., expected \mbox{long-term} reward starting from $(x,y)$ (right).}
    \label{fig:seq}
    \vspace{-0.3cm}
\end{figure}

\begin{figure}
    \centering
    \includegraphics[width=\columnwidth]{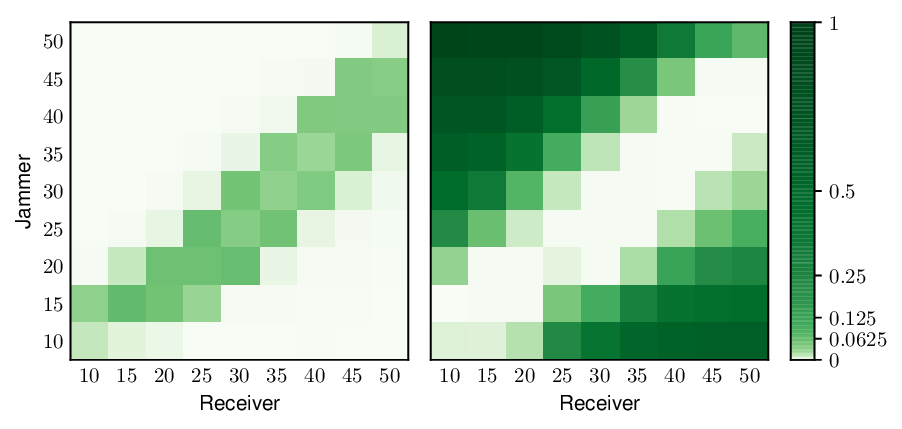}
    \caption{Game $\mathcal{G}_2$, simultaneous game with complete but imperfect information. Joint probability for R and J to find themselves in position $(x,y)$ (left). State value function of R, i.e., expected \mbox{long-term} reward starting from state $(x,y)$ (right). }
    \label{fig:pos}
\end{figure}

The logarithmic heatmaps of Figs.~\ref{fig:seq} and~\ref{fig:pos} show, on the left, the joint probability of finding R and J at positions $x$ and $y$, respectively, and, on the right, R's state action value function~\eqref{eq:value} normalized in $[0,1]$, for complete information games with sequential ($\mathcal{G}_1$) and simultaneous moves ($\mathcal{G}_2$), respectively. The sequential game $\mathcal{G}_1$ presents a simple equilibrium: R ``bounces'' between positions of index $1$ and $2$, while J can only follow it reactively (see Fig.~\ref{fig:seq}). This means that $50\%$ of the time R is prevented from transmitting (when it is caught by J), but the remaining $50\%$ of the time it is allowed to transmit close to the AP. This translates into a payoff of $13/144$ for R with the chosen distances. In this case, the learned state-action value function is a coarse approximation of the real one because the players learn the optimal policy quickly without needing to explore the environment.
Now, we consider $\mathcal{G}_2$, where, instead, the players move simultaneously. As can be seen from the state-action value function of Fig.~\ref{fig:pos} (here approximated with a higher resolution), R's favorite states are the ones where it is far from J, i.e., when J is close to $L$ and R is close to $M$, or vice versa. However, since J knows R's position, it forces R to visit the complementary states to its favorites (figure on the left). This confirms the very low average payoff reached by R in these conditions: concerning the sequential game analyzed previously, the payoff is about halved (Fig.~\ref{fig:convergence}).

\begin{figure}
    \centering
    \includegraphics[width=\columnwidth]{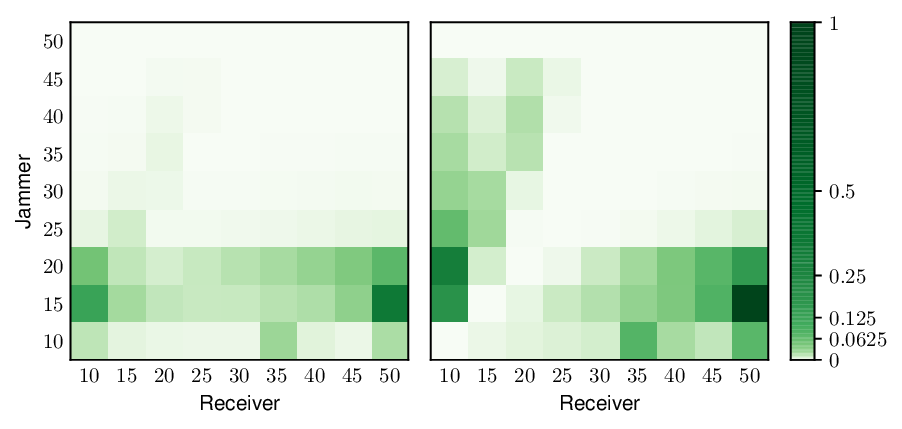}
    \caption{Game $\mathcal{G}_3$, simultaneous game with imperfect and incomplete information. Joint probability for R and J to find themselves in position $(x,y)$ (left). Expected instantaneous reward, weighted by the joint position probability (right). }
    \label{fig:no_pos}
    \vspace{-0.3cm}
\end{figure}

Finally, in Fig.~\ref{fig:no_pos}, the game $\mathcal{G}_3$ with simultaneous moves and incomplete information is considered. The long-term value of a state cannot be represented anymore, because the \revisionthree{state is no longer defined over the joint position space}, and the $Q$~value is only a function of~$x$, and not of~$y$. Therefore, together with the joint probability of the position pair $(x,y)$, the expected instantaneous reward is plotted, i.e., R's payoff in state~$x$ while J is in state~$y$, weighted by the probability that R is indeed in~$x$. Since J does not know R's position, it more often chooses to be close to the AP to prevent R from being there too, in accordance with Theorem~\ref{rem:2}. As \revisionthree{shown by} the heatmap on the left, J lingers mostly around $15$-$20$ meters from the AP, a result consistent with J's optimal position according to the outcome of the static game. Consequently, R learns that it must stay away from J, going toward the opposite side of the geometry, or very close to the AP, which gives a very similar payoff, with almost equal probability. This is consistent with R's optimal strategy of playing either $L$ or~$M$.

\begin{figure*}[tb]
    \centering
    \subfloat[\label{fig:lj_1}Learning jammer, $S=1$]{\includegraphics[width=0.33\textwidth]{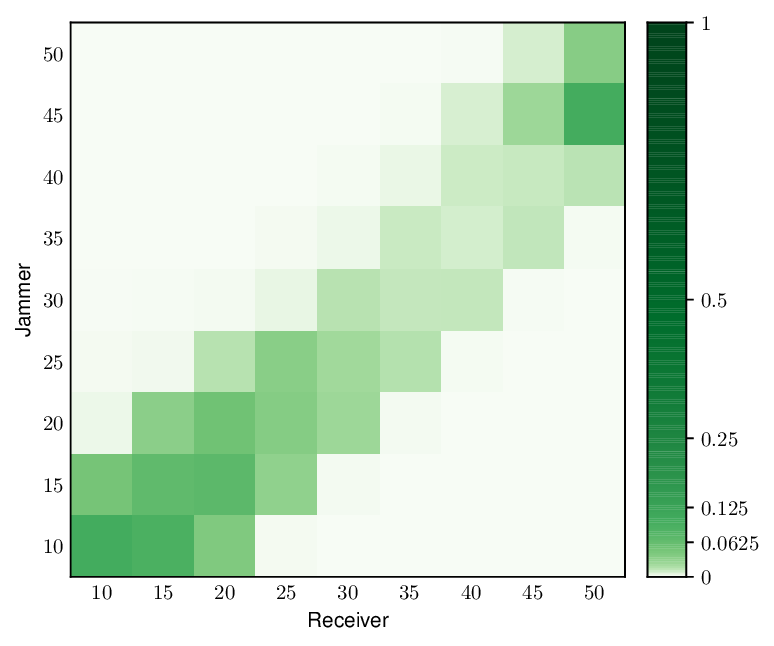}}
    \subfloat[\label{fig:gj_1}Greedy jammer, $S=1$]{\includegraphics[width=0.33\textwidth]{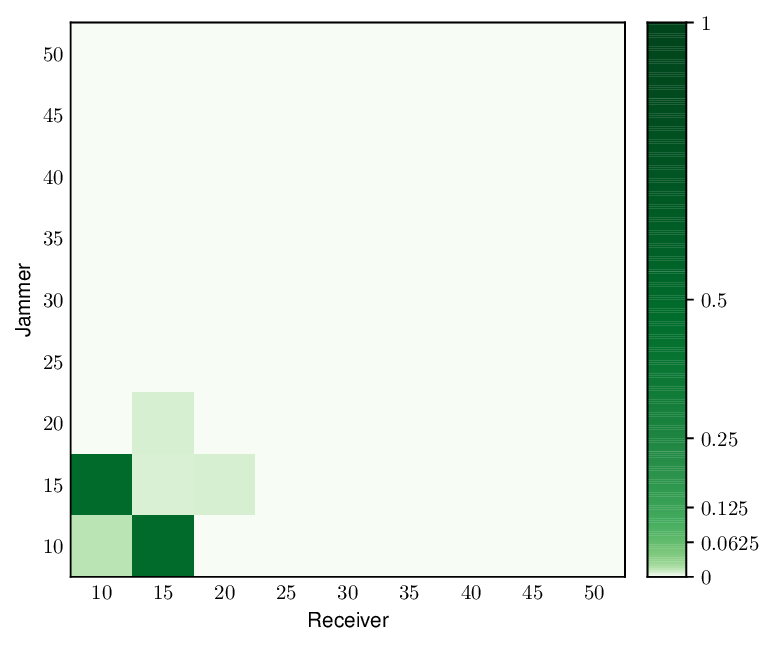}}
    \subfloat[\label{fig:mj_1}Mixed strategy jammer, $S=1$]{\includegraphics[width=0.33\textwidth]{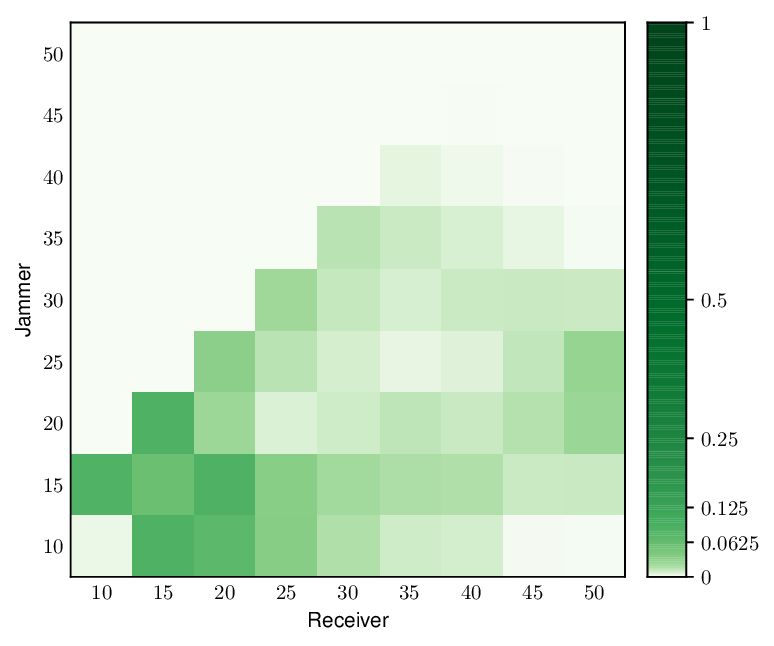}}\\
    \subfloat[\label{fig:lj_2}Learning jammer, $S=2$]{\includegraphics[width=0.33\textwidth]{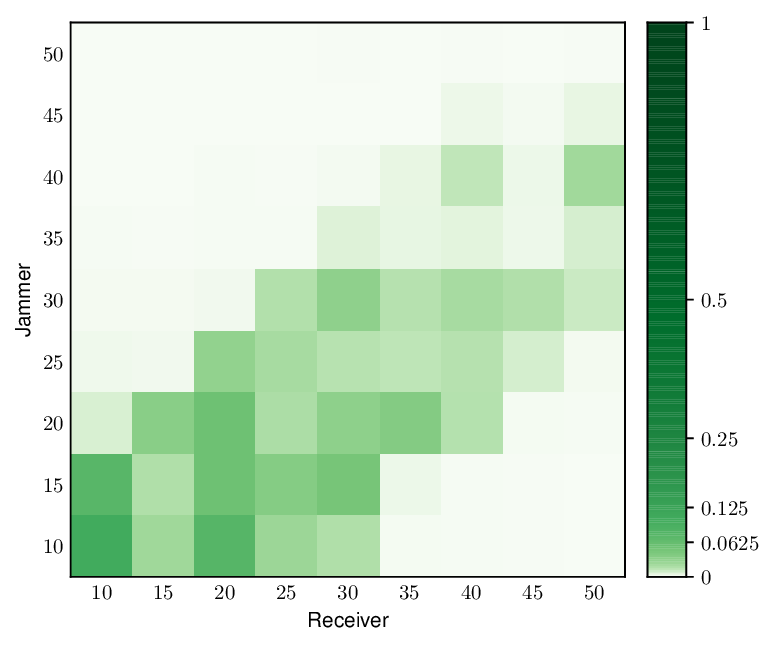}}
    \subfloat[\label{fig:gj_2}Greedy jammer, $S=2$]{\includegraphics[width=0.33\textwidth]{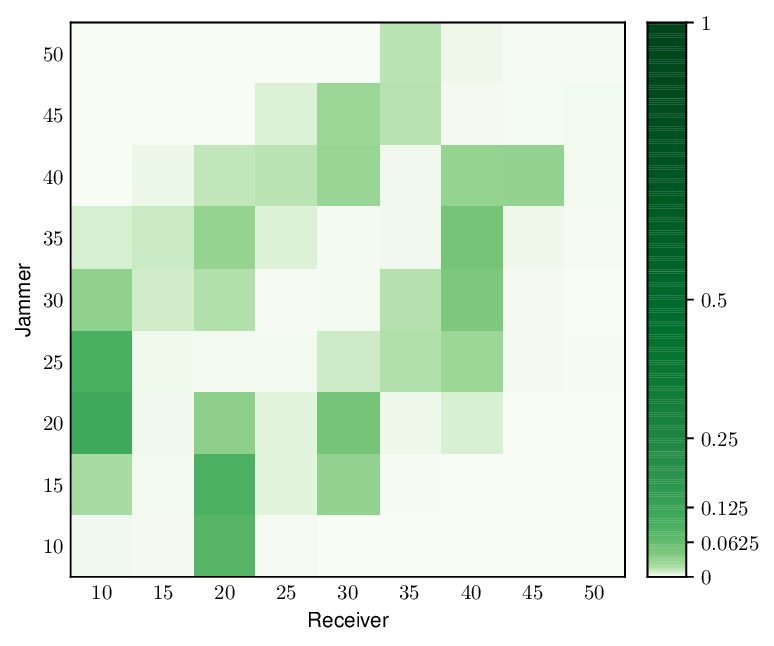}}
    \subfloat[\label{fig:mj_2}Mixed strategy jammer, $S=2$]{\includegraphics[width=0.33\textwidth]{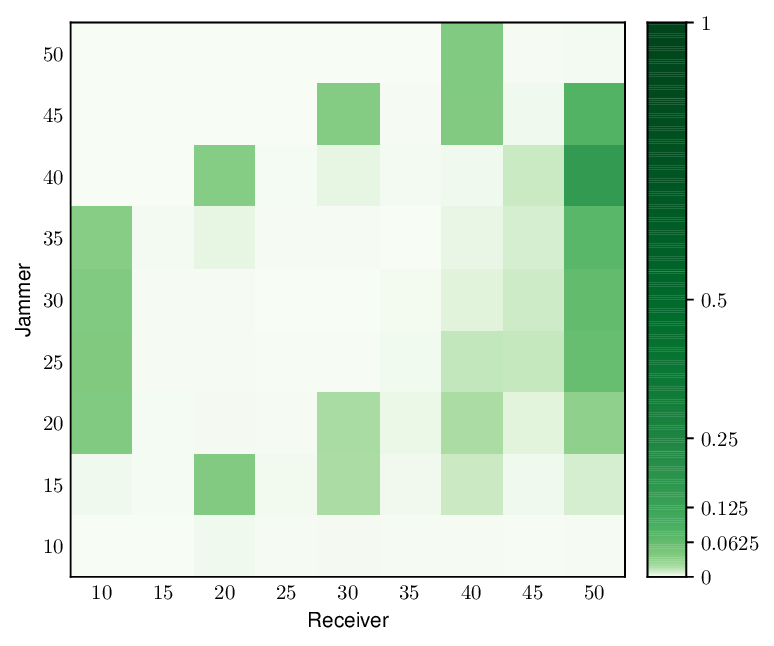}}
    \caption{Deep RL-based policies for game $\mathcal{G}_2$ under different jammer behaviors.}
    \label{fig:DRL_policies}
\end{figure*}
\subsection{Deep RL-based policies for game $\mathcal{G}_2$}
In this section, \revisionthree{DRL is} applied to game $\mathcal{G}_2$ in various contexts. Specifically, we i) explore what happens when $S=2$ and ii) verify the learning behavior of player R when J follows greedy policies driven by Theorem~\ref{rem:1}. Fig.~\ref{fig:lj_1} shows the joint policy of R and J when J is a learning player and $S=1$. It is the same result reported in Fig.~\ref{fig:pos} but using the DRL approach: since the policies are very similar, the DRL approach is validated. J plays rationally and can follow R closely wherever it moves; conversely, R tries to escape J throughout the allowed span. \revisiontwo{The average payoff for configuration (a) is $\bar u_{\rm R}^{\rm{(a)}} = 0.06 \pm 0.01$.} However, when $S=2$ (Fig.~\ref{fig:lj_2}), R can threaten J by moving one or two positions, thus being able to keep a higher distance from J on average. In particular, we can observe how, when J is close to the AP, R can directly ``jump'' from position $x_1 = 10$~m to $x_2 = 20$~m. Using $S=2$ to increase \revisionthree{the moves available to the players} is beneficial to R because it is harder for J to predict R's behavior: the average payoff increases to $\bar u_{\rm R}^{\rm{(d)}} = 0.12 \pm 0.01$, i.e., a \revisionthree{twofold gain over} configuration (a). A \emph{greedy jammer} is tested in Fig.~\ref{fig:gj_1} with these rules: a) if J and R are in the same position, J keeps that position; b) if they are not, J moves toward R. In this case, an RL-based receiver easily learns that J has consistently reactive behavior and can move between the two closest positions to the AP: J is always one step \revisionthree{behind} R, and it is never convenient for R to move farther because J will follow closely. \revisiontwo{With a non-learning jammer, even the case $S=1$ yields a higher payoff of $\bar u_{\rm R}^{\rm{(b)}} = 0.18 \pm 0.01$.}
Fig.~\ref{fig:gj_2} shows the game associated with $S=2$: R can effectively exploit the possibility to ``jump'' two cells, namely, to move faster than J. 
\revision{Since the closest position to the AP is indexed 1 (odd index) and moving faster is beneficial ($S=2$), R will find \revisionthree{itself} most often in an odd-indexed position, as a result of summing the preferred odd-indexed position 1 with an even move of $S=2$ positions.} \revisiontwo{For this configuration we obtain a gain of about $3.7\times$ with respect to $S=1$, yielding a payoff $\bar u_{\rm R}^{\rm{(e)}} = 0.67 \pm 0.02$.}
Finally, a \emph{mixed strategy jammer} is also implemented with the following rules: a) if J and R are in the same position, J keeps that position or moves backward with uniform probability; b) if J is closer to the AP than R, it keeps the position or moves forward with uniform probability; c) if R is closer to the AP, J moves backward. From Fig.~\ref{fig:mj_1}, it can be seen that some randomness in the behavior of J constrains R to remain farther away from the AP, on average. However, this is still a suboptimal policy for J as it is not able to always closely follow R. \revisiontwo{The payoff for this configuration is $\bar u_{\rm R}^{\rm{(c)}} = 0.17 \pm 0.01$, which is similar to the payoff of configuration (b) although the strategy played is completely different.} The corresponding situation with $S=2$ is shown in Fig.~\ref{fig:mj_2}. R can still benefit from moving faster than J, but the same reasoning holds: the increased randomness in J's behavior keeps R farther from the AP on average, \revision{but still preferring odd-indexed positions,} as per the result of Fig.~\ref{fig:gj_2}. \revisiontwo{The payoff is indeed similar, with a value of $\bar u_{\rm R}^{\rm{(f)}} = 0.68 \pm 0.02$ but the distribution is different: here, the majority of the payoff mass is obtained close to the extremities $L$ and $M$, while in configuration (e) $M$ is never visited. The gain over configuration (c) with $S=1$ is $4\times$.}
\begin{figure}
\centering
\resizebox{\columnwidth}{!}{\input{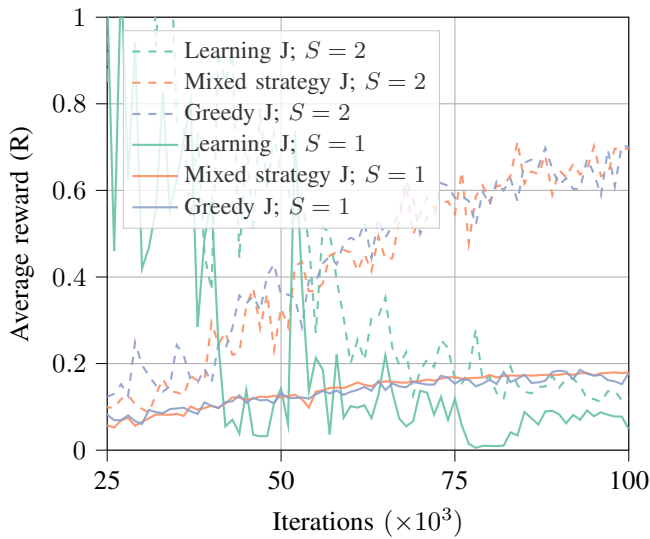}}
    \caption{Receiver's average reward of the DRL approach for game $\mathcal{G}_2$ in the six considered scenarios. The result is filtered with a moving average of width $W=500$ samples.}
    \label{fig:convergence_deep}
    \vspace{-0.3cm}
\end{figure}

In Fig.~\ref{fig:convergence_deep}, R's payoffs using the DDQN approach under the six different scenarios involving $\mathcal{G}_2$ are shown \revisiontwo{vs. the iteration index during learning. We plot a single run as an example, as other runs were similar.} As expected, fewer iterations are required to reach a stable solution \revisionthree{than with} the classic tabular approach, due to the ability of deep learning to generalize for a large state-action space (in particular when $S=2$). This plot also confirms \revisiontwo{what previously observed on the average payoffs for the different scenarios. Specifically,} (i) $S=2$ is beneficial to R regardless of J's policy; (ii) although a mixed strategy J induces R to change strategy, the expected payoff is very similar to the case of greedy J; and (iii) a learning jammer can effectively improve its performance, especially when $S=2$.

\noindent\textbf{Convergence speed.}
As shown above, even with the deep learning approach, the number of iterations (i.e., action-observation \revisionthree{pairs}) required to find a stable result is beyond the possibilities of real moving agents. This is because of the time required to physically perform the movement action. However, training can be done offline, and a pre-trained policy for a specific technology (e.g., IEEE 802.11p) can be stored at each player. Agents can then \revisionthree{adapt the policy online} to environment-specific behaviors (e.g., the shadowing pattern caused by reflections on buildings) while having a good average initial guess.
\subsection{Dependency on network parameters}
\label{sec:res-net-params}

\begin{figure}
    \centering
    \resizebox{\columnwidth}{!}{\input{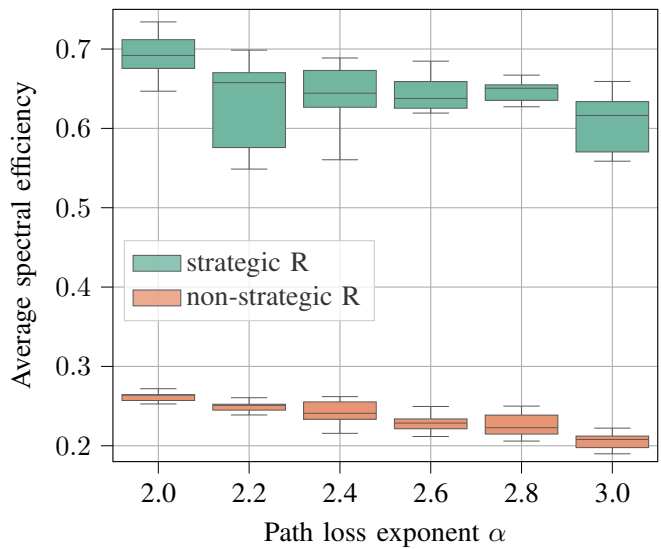}}
    \caption{\revision{Comparison between a strategic and a non-strategic receiver in terms of spectral efficiency when a strategic jammer is present. Noise floor $N_0=-174$~dBm/Hz; bandwidth $B=20$~Mhz; log-normal shadowing with variance $\sigma^2=2.5$~dB; $L=10$~m, $M=570$~m, $N=15$, $S=2$.}}
    \label{fig:strategic}
    \vspace{-0.3cm}
\end{figure}

In this paragraph, the advantage of being strategic in the dynamic setting is discussed considering a scenario that includes noise and shadows. Specifically, the reward given to the receiver is now the spectral efficiency, that is,
\begin{equation}
    u_{\rm R}=-u_{\rm J}=\log_2\left(1+\frac{g_{\rm R} P_{\rm tx}}{g_{\rm J} P_{\rm J} + BN_0}\right),
\end{equation}
where, given a normal random variable $\mathcal{N}(0, \sigma^2)$, the receiver's and the jammer's channel gains are retrieved as
\begin{equation}
    \begin{aligned}
        (g_{\rm R})_{\rm dB} = - (47.86 + 10\alpha\log_{10}(x) + \mathcal{N}) \quad \text{and}\\
        (g_{\rm J})_{\rm dB} = - (47.86 + 10\alpha\log_{10}(\lvert x - y \rvert) + \mathcal{N}),
    \end{aligned}
\end{equation}
respectively (the numeric coefficient is specific for vehicular communications; see the log-distance path loss model in~\cite{etsi44103}). \revision{In Fig.~\ref{fig:strategic}}, a comparison is shown between the following approaches.
\begin{description}
    \item[Strategic R] using the DDQN to learn the policy;
    \item[Random R] choosing an action randomly with uniform distribution~\cite{garnaev2020jamming};
    \item[Greedy R] following a deterministic smart policy: if J is in the same position as R, choose any action with uniform probability; if J is closer than R to the AP, move right if possible (i.e., away from J); if J is farther than R to the AP, move left if possible (i.e., towards the AP).
\end{description}

The resulting spectral efficiency obtained \revisionthree{over $10$ independent runs of} game $\mathcal{G}_2$ with $L=10$~m, $M=570$~m, $N=15$, and $S=2$ while varying the path loss exponent in $[2,3]$, i.e., \revisionthree{from LoS to non-LoS conditions}, is depicted, together with a linear fit. As can be seen, being strategic allows a gain of about $2.5 \times$ \revisionthree{that of} the random approach, where the jammer easily performs its task of disturbing the communication. \revisiontwo{For a strategic R, the spectral efficiency fit decreases from $\bar u_{\rm R}^{\rm s}(\alpha=2)=6.66 \times 10^{-1}$ ($10\%-90\%$ confidence interval $[6.44, 6.87]\times 10^{-1}$) to $\bar u_{\rm R}^{\rm s}(\alpha=3)=5.98 \times 10^{-1}$ (c.i. $[3.91, 6.63] \times 10^{-1}$). For the random R, instead, the spectral efficiency goes from $\bar u_{\rm R}^{\rm r}(\alpha=2)=2.39$ (c.i. $[2.27, 2.55]\times 10^{-1}$) to $\bar u_{\rm R}^{\rm r}(\alpha=3)=1.92 \times 10^{-1}$ (c.i. $[1.75, 2.04]\times 10^{-1}$).} The greedy approach is the least performing: While it might seem odd that acting randomly results in higher spectral efficiency, this can be explained by the fact that a learning jammer can better predict deterministic actions than random actions. As a consequence, a learning jammer will be able to follow closely and even trap a receiver behaving deterministically. \revisiontwo{In this latter case we have $\bar u_{\rm R}^{\rm g}(\alpha=2)=1.52 \times 10^{-2}$ (c.i. $[1.37, 1.86] \times 10^{-2}$) to $\bar u_{\rm R}^{\rm g}(\alpha=3)=1.31 \times 10^{-2}$ (c.i. $[0.80, 2.17]\times 10^{-2}$). Overall, we notice that the variance is larger for the strategic R as a consequence of the fact that both players are being strategic and adversarial, trying to counteract the opponents' moves.}

\revision{Note that there is a very weak dependency on the path loss exponent: the spectral efficiency decreases slightly (linearly) with increasing $\alpha$, even in the presence of noise and slow fading. This happens mainly for two reasons. First, the jammer's power is much higher than the noise power, which makes the presence of the noise negligible when the jammer is close to the receiver. In a dynamic setting, a learning jammer will always be close to the receiver because of its ability to move. Second, even if the increase in the path loss exponent heavily affects farther positions from the AP, a learning receiver will just stay closer to it to avoid such positions.}

\section{Conclusions and Future Work}
\label{sec:conclusion}
We discussed a jamming scenario in which the nodes involved move during subsequent rounds. This represents an original extension of the existing literature that may lead to interesting findings. We considered \revisiontwo{a static and three dynamic setups} with variations related to the timing of the moves of the players and the information available to them. We found that the dynamic games obtain interesting results differing significantly from the characterization \revisiontwo{of the static game}. Nevertheless, some general principles \revisiontwo{of the mathematical analysis} can be \revisiontwo{extended} and are confirmed by our \revisiontwo{simulation campaign.}

More complex and distributed ways to train the agents can be designed. For example, \cite{mowla2020afrl} proposed a federated RL for a similar problem of avoiding jamming but without involving mobility, and, therefore, an extension along these lines can be interesting. More generally, even though our approach based on (deep) Q-learning with $\epsilon$-greedy \mbox{exploration-exploitation} has \revisionthree{proved} effective, we conjecture that some original developments are possible for a dedicated technique for this specific problem. Especially, we remark that the reinforcement learning approach \revisionthree{in} the scenario with incomplete information just tries to estimate the optimal policy by blind iterations. However, mutual assumptions of the rationality of the players would suggest that some strategic choices can be anticipated, and also a joint estimate of multiple environmental parameters (despite the tremendous growth in the search space) would be more effective in reaching the equilibrium.
At the same time, different mobility scenarios and patterns can also be adopted. \revisionthree{These extensions constitute promising directions for future work}.

\section*{Acknowledgments}
\noindent
This work was supported by the European Commission through the Italian Ministry of University and Research under the NRRP RESTART Program -- Mission 4, Component 2, Investment 1.3, theme 14 ``Telecommunications of the future''.

\appendix
\section{Proof of Theorem \ref{theo:1}}
 \label{app:1}

The theorem follows directly from Parthasarathy's theorem~\cite{parthasarathy1970games}.
If J believes that R is choosing a pure strategy, i.e., a location $x$ with probability equal to $1$, its best response would be to choose $y=x$, since, in that case, the value of the game would be $0$, which is the minimum possible value. Being rational, R anticipates this and knows that it would be better off with any unilateral deviation. As a consequence, R must necessarily play a mixed strategy at the NE.
\qed

\revision{
\section{Proof of Theorem \ref{theo:thresh}}
\label{app:1b}
The value of the game saturates as R goes further from the AP, which mathematically means that $\frac{\partial u_{\rm R}(x, y)}{\partial x}$ decreases uniformly for all $y$. Thus, we can find $\tilde{M}$ as the supremum of the valid locations where moving further away from the jammer can still yield a non-negative marginal utility:
\begin{equation*}
    \tilde{M} = \sup \Big\{ x \in [L,M) \mid \exists \,y \in [L,M] \text{ s.t. } \frac{\partial u_{\rm R}(x, y)}{\partial x} \geq 0 \Big\}
\end{equation*}
By the uniform convergence of $u_{\rm R}$, this supremum is finite, being either $\tilde{M} = M$ or $\tilde{M} < M$. In the former case, the theorem is trivially true as there is no further position beyond $M$ where R can go.
In the latter case, the marginal utility satisfies $\frac{\partial U_R(x, y)}{\partial x} < 0$ for $x \in [\tilde{M},M]$ and $y \in [L,M]$. Integrating this strict inequality over the interval $[\tilde{M}, x]$ yields $U_{\rm R}(\tilde{M}, y) > U_{\rm R}(x, y)$ for all $y \in [L, M]$, proving that any $x \in (\tilde{M},M]$ is strictly dominated by $\tilde{M}$. 
\qed
}

\section{Proof of Theorem \ref{theo:NEgen} }
\label{app:2}

The theorem corresponds to showing the existence and uniqueness of a joint strategy where both players play a best response to each other's move. According to Theorem \ref{theo:1}, the strategy of R must be strictly mixed.
Thus, the theorem can be proven through three steps: first, we show the existence of a particular value $\revision{y^\star}$, which lies inside the interval $[L,W]$, i.e., $L<\revision{y^\star}<W$, with $W\le \tilde M$, and such that choosing $y = \revision{y^\star}$ makes the best response of R as indifferently choosing either $x=L$ or $x= W$, or a probabilistic mixture thereof. 

\revision{The existence of the point $W \le \tilde M$ is ensured by Theorem~\ref{theo:thresh}: since $\tilde M$ is the point beyond which R's strategies are strictly dominated, any BR must be a point $W\le \tilde M$.}

Secondly, we show that player R can properly choose a mixed strategy that has support in $x=L$ and $x=W$ only, with the respective probabilities within the mixture such that the best response of player J is to choose $y=\revision{y^\star}$.
The third and last point is to conclude that the proposed pair of strategies, consisting of a pair of mutual best responses for the players and therefore a NE, is also unique. 

The first step immediately follows from $u_{\rm R}(L,y)$ being monotonically increasing in $y$, while $u_{\rm R}(W,y)$ is monotonically decreasing, and additionally, they are respectively zero when $y=L$ or $y=W$. This implies that there is a unique solution in the unknown $y$ to condition $u_{\rm R}(L,y) = u_{\rm R}(W,y)$, which we call $\revision{y^\star}$.

For the second point, we remark that for any $y \in [L,W]$ as the choice of player J, the utility of player R is decreasing or increasing in $x$ for $x \in [L,y]$ or $x \in [y,W]$, respectively. Thus, the utility of R is maximized by $x=L$ or $x=W$, or possibly both. The last case occurs only for $y=\revision{y^\star}$. Since according to Theorem \ref{theo:1}, the NE cannot be in a pure strategy of R, the only option for an NE is that both $x=L$ and $x=W$ are best responses, which implies that any mixture of these two pure strategies is also a best response, and therefore we must choose $y=\revision{y^\star}$.
This shows that $u_{\rm R}(L,\revision{y^\star}) = u_{\rm R}(W,\revision{y^\star}) = u_{\rm R}^\star$ is the \emph{value} of the game at the NE. 


\revisionthree{Conversely, consider R mixing between $x=L$ and $x=W$ with probabilities $p$ and $1-p$, respectively. By continuity and the opposite monotonic behavior of $u_{\rm R}(L,y)$ and $u_{\rm R}(W,y)$, there exists $p^\star\in(0,1)$ such that $y^\star$ minimizes the convex objective
$p^\star u_{\rm R}(L,y)+(1-p^\star)u_{\rm R}(W,y)$.
Hence, $y^\star$ is a best response of J to this mixed strategy of R, while L and W are both best responses of R to $y^\star$.}

Thus, we found a pair of strategies consisting of mutual best responses, which is therefore an NE. The uniqueness of this NE formally follows from this being a game on the unit square with continuous value, as per \cite{glicksberg20169}.
\qed
 
\section{Proof of Theorem \ref{theo:NE} }
\label{app:3}

This is a specific instance of Theorem \ref{theo:NEgen}, where $\revision{y^\star}$ and $p^\star$ can be computed in closed form.
As for the general case, the support of R's best response at the NE (i.e., the set of pure strategies played with non-zero probability) only contains $x=L$ and $x=M$. 
This pair of strategies satisfies the characterizing property of a NE according to the \emph{indifference criterion}~\cite{tadelis}. 
\revision{By imposing $u_{\rm R}( L, \revision{y^\star} )  = u_{\rm R}( M, \revision{y^\star})$ from Eq.~\eqref{eq:valfin}, we can get the value of $y^\star$ as
\begin{align}
    \frac{\lvert L-y^\star\rvert^\alpha}{L^\alpha} &= \frac{\lvert M-y^\star\rvert^\alpha}{M^\alpha}\nonumber\\
    M(y^\star -L) &= L(M-y^\star)\nonumber\\
    y^\star &= \frac{2LM}{L+M},
\end{align}
where the absolute values yield only one possible case as we know that $L<y^\star<M$. It is immediate that plugging back $y = \revision{y^\star}$ in Eq. (\ref{eq:valfin}) we get
}
\revisionthree{\begin{equation}
u_{\rm R}( L, \revision{y^\star} )  = u_{\rm R}( M, \revision{y^\star})  = \left(\frac{M-L}{L+M}\right)^\alpha
.
\end{equation}}
Conversely, for $L < x < M$ we have $u_{\rm R}( x, \revision{y^\star} ) =  \big[(1 - \revision{y^\star} / x )\sgn(x-\revision{y^\star})\big]^\alpha$, where $\sgn$ is the sign function, and this payoff is always strictly less than the value at $L$ or $M$.

To conclude, we need to remark that the respective probabilities of $x=L$ or $x=M$ at the NE must be $p^\star = L/(L+M)$ and $1-p^\star$. This is proven by showing that, if R chooses this specific mixed strategy, then the best response of J is $\revision{y^\star}$. 
\qed

\section{Proof of Theorem \ref{theo:Stack}}
\label{app:4}
This follows from Theorem \ref{theo:NEgen}, based on the minimaximization (minimax) principle \cite{du1995minimax}. 

Since the setup is adversarial, the minimax and maximin in mixed strategies coincide with the payoff at the NE. Also, since according to Theorem \ref{theo:NEgen}, the NE of $\mathcal{G}_0$ implies that J plays a pure strategy, this player can also choose this very same strategy if moving first. Then, R is free to choose $x=L$ or $x=M$, or even a mixture of them, and the value will be the same. 

If R is the leader instead, the situation changes in that the NE is achieved only if R is allowed to choose a mixed strategy (in which case R should choose that of the NE). If R is bounded to a pure strategy instead, whatever its choice, J will always choose the same position, thus achieving zero value. Hence, if a rational player J moves after knowing the position of the receiver, player R cannot maximize the worst-case value (maximin) beyond $0$, which is the guaranteed outcome for J (minimax).
\qed

\section{Proof of Theorem \ref{rem:1}}
\label{app:5}
This is a consequence of Theorem \ref{theo:NE} and the discussion thereof. Whenever J is in the same position as R, the value of the game goes to $0$. For any other choice of position, the value is positive, as per \eqref{eq:valfin}. Also, R's payoff strictly decreases as $y$ approaches $x$, thus, J has no incentive to move away from R. In other words, moving away always gets lower payoff than staying still. As a side note, moving in the direction of R is often better, but border effects may not guarantee the dominance to be strict.  \qed

\section{Proof of Theorem \ref{rem:2}}
\label{app:6}
In the static game $\mathcal{G}_0$, J chooses $\revision{y^\star}$ at the NE. However, this is a best response to a mixed strategy where R randomizes between the two extreme points $x=L$ and $x=M$. In dynamic problems, even when the position of the opponent is unknown (such as $\mathcal{G}_3$), it is at least certain that R's position corresponds to a single value. 
Since it was proven that $\revision{y^\star}$ falls within $L$ and $2L$, the latter being the limit if $M$ becomes very large, regardless of the actual position of R, J can upper bound the utility of R below $1$ by choosing any dynamic strategy that keeps J's position between $L$ and $2L$. \qed



\bibliographystyle{elsarticle-num}
\balance
\bibliography{IEEEabrv, jambad}

\end{document}